\documentclass[letter]{aa} 
\usepackage{graphicx}
\usepackage{txfonts}
\begin{document}
   \title{The masses of hot subdwarfs }

   \subtitle{}

   \author{X. Zhang
          \inst{1,2}
          \and
          X. Chen\inst{1}
          \and
          Z. Han\inst{1}
          }


   \institute{National Astronomical Observatories/Yunnan Observatory, the
              Chinese Academy of Sciences, Kunming 650011, PR China\\
              \email{X.F.Zhang@live.com}
         \and
             Graduate School of the Chinese Academy of Sciences\\
             }

   \date{Received  ,  ; accepted  ,  }


  \abstract
  {Masses are a fundamental parameter, but they are not well known for most hot
   subdwarfs. In general, the mass of a hot subdwarf is derived with asteroseismology or dynamical methods,
   for which it is often difficult to obtain the necessary data from observations.}
   {We intend to find an approach to deriving the masses of hot subdwarfs from observational data in the literature.}
   {We presented full evolutionary calculations
    for hot subdwarfs in a wide mass range (0.33 $M_\odot$ to 1.4 $M_\odot$)
    for a Population I metallicity of $Z$=0.02, and obtained a relation between $M_{\rm p}$ and $\log (\frac{T_{\rm eff}^4}{g})$, where
    $M_{\rm p}$, $T_{\rm eff}$, and $g$ are the most probable
    mass, effective temperature, and gravity. This relation is used to study the masses of some observed hot subdwarfs.}
   {We proposed a method of determining the masses of hot subdwarfs.
   Using this method, we studied the masses of hot subdwarfs from the ESO supernova Ia progenitor survey and Hamburg quasar survey.
   The study shows that most of subdwarf B stars have masses between 0.42 and 0.54 $M_\odot$,
     whilst most sdO stars are in the range 0.40 $\sim$ 0.55 $M_\odot$.
    Comparing our study to the theoretical mass distributions of Han et al. (2003),
    we found that sdO stars with mass less than $\sim$ 0.5 $M_\odot$ may evolve from sdB stars,
    whilst most high-mass($>$ 0.5 $M_\odot$) sdO stars result from mergers directly.}
   {}


   \maketitle
%

\section{Introduction}

   Hot subdwarfs are generally classfied into three types by their
   spectra. These are subdwarf B (sdB, with a surface effective temperature, $T_{\rm eff}$,
   in a range from 20,000 to 40,000K, with H-Balmer
   absorption lines wider than in normal B stars), subdwarf O (sdO, $T_{\rm eff}$ ranging from 40,000K to 80,000K
   with strong He absorption lines), and subdwarf OB (sdOB, a
   transition between O and B, Moehler et al. 1990; Heber 2009). These objects are located below the upper main
   sequence on the Hertzsprung-Russell diagram(HRD) and are also
   known as extreme horizontal branch (EHB) stars from the view of their
   evolutionary stages; i.e., they are believed to be core He-burning objects
   with extremely thin hydrogen envelopes ($<$ 0.02 $M_\odot$).

   Hot subdwarfs are an important population in several respects. For example, pulsating
   sdB stars are standard candles in distance determination (Kilkenny et al. 1999).
   Likewise, close binaries composed of an sdB star and a massive white dwarf (WD)
   are qualified as supernova Ia progenitors (Maxted et al. 2000).
   Moreover, hot subdwarfs are an important source of far-UV light in the
   galaxy, and they are successfully used to explain the UV-upturn in
   elliptical galaxies (Kilkenny et al. 1997; Han et al. 2007).

   Several scenarios have been proposed
   to explain the formation of these objects, i.e. strong mass loss for a star on the
   red giant branch (Dorman et al. 1993, D'Cruz et al. 1996),
   mass transfer in close binary systems (Mengel et al. 1976), and the coalescence
   of two helium white dwarfs (He-WDs) (Iben 1990, Webbink 1984). Recent radial velocity (RV) surveys reveal that
   two thirds of all sdB stars reside in close binaries (Maxted et al. 2001). Han et al. (2002,
   2003) propose a binary model for the formation of hot subdwarf stars, in which three
   channels (stable Roche lobe overflow (RLOF), common envelope (CE) ejection, and merging of two
   He-WDs) are included. (For details see Han et al. 2002, 2003.) This binary model successfully explained most of the observational
   characteristics of hot subdwarfs, and is now widely used in the study of hot subdwarf stars, e.g. O'Toole, Heber \& Benjamin,
   2004.

   Mass is a fundamental parameter of stars, but it is very uncertain for most hot subdwarfs,
   which are {\it assumed} to be around 0.5 $M_\odot$
   (Heber 1986, Saffer et al. 1994). Sometimes we need a more precise mass
   to study these objects further. In recent years, more than 2300
   objects have been included in a hand hot subdwarf database ({\O}stensen 2004),
   among these more than 200 hot subdwarfs have been studied in detail for atmospheric
   parameters (e.g. Saffer et al. 1994; Maxted et al. 2001;
   Edelmann et al. 2003, Lisker et al. 2005, Stroeer et al. 2007).
   However, only a few hot subdwarfs have well-defined masses
   from asteroseismology, e.g. PG 0014+067 (Brassard et al. 2001),
   PG 1219+534 (Charpinet et al. 2005a), PG 1325+101(Charpinet et al.
   2006), and Feige 48 (Charpinet et al. 2005b). In addition, Wood \& Saffer
   (1999) show an sdB star HW Vir (PG 1241-082), which is in a double-lined
   eclipsing binary, to be $0.48 \pm 0.09 M_\odot$ from dynamical methods.
   For other hot subdwarfs, which are not pulsating stars
   or are in binary systems without detailed light or radial-velocity curves, masses
   are derived from theoretical evolutionary tracks (Dorman et al.
   1993), e.g., choosing the track closest to the data point on the $T_{\rm eff}- \rm
   log(\it g)$ diagram, where $T_{\rm eff}$ and $g$ are effective temperature and surface
   gravity, respectively. The commonly used evolution library of hot subdwarfs is that of Dorman et al.
   (1993), with masses from 0.4 to 0.5 $M_\odot$. However, Han et al.(2002, 2003)
   show that the masses of hot subdwarfs can be in a much wider
   range; i.e., from 0.3 $M_\odot$ to more than 0.7 $M_\odot$. Thus, it is
   necessary to extend the evolution library of hot subdwarfs or
   find a simple method of estimating the masses of hot subdwarfs.

   In this letter, we carry out full evolutionary calculations
   for hot subdwarfs and obtain an approach to determining the masses of hot
   subdwarfs. Using this approach, we study the masses and mass distributions of observed hot
   subdwarfs.


\section{The model}

   The stellar evolution code used was originally
   developed by Eggleton (1971, 1972, 1973). The code has been updated
   with the latest input physics over the past three decades as
   described by Han, Podsiadlowski \& Eggleton (1994, hereafter HPE)
   and by Pols et al. (1995, 1998). We set the ratio of mixing length to local
   pressure scale height, $\alpha = l/H_{\rm p}$, to 2.0 and the
   convective overshooting parameter, $\delta_{\rm OV}$, to 0.12
   (Pols et al. 1997; Schr$\ddot{o}$der et al. 1997). The opacity
   tables for various metallicities are compiled by Chen \& Tout (2007) from
   Iglesias \& Rogers (1996) and Alexander \& Ferguson (1994). The
   initial hydrogen mass fraction, $X$, is obtained by, $X=0.76-3.0Z$, where $Z$ is the metallicity (Pols et al.
   1998).

\begin{figure}
    \centering
     \includegraphics [angle=0,scale=0.3]{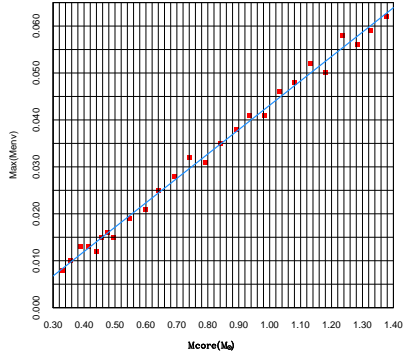}
     \caption{The maximum envelope masses for different core masses
      in our hot subdwarf models (squares). The solid line is
      a linear fitting of the square points, with errors less than 0.001$M_\odot$.}
      \label{fig1}
   \end{figure}

   We constructed a series of zero-age extreme horizontal branch
   models (ZAEHB, i.e. helium begins stably burning
   in the core) of Pop I ($Z$=0.02) and carried out their evolutions in detail. The
   model grid contains a wide mass range; i.e., the core mass ranges
   from 0.33 to 1.4 $M_\odot$, and the envelope mass changes, by a
   step of about 0.005 $M_\odot$, from
   zero to the maximum envelope mass (see Fig.\ref{fig1}), beyond which the constructed object
   is not a hot subdwarf any more (i.e. $T_{\rm eff}<$ 20,000K or $\rm log(\it g)<$
   4.5).

   As an example, we show the evolutionary tracks of hot subdwarfs with
   a core mass of 0.475 $M_\odot$ in Fig.2. The envelope masses, $M_{\rm env}$, are
   0.000, 0.001, 0.005, 0.010, and 0.015 $M_\odot$.
   TAEHB is the termination-age EHB, i.e., at the core helium exhaustion (Dorman et al. 1993).
   The region between ZAEHB and TAEHB is known as the main sequence.
   The age difference between two neighbouring crosses is 10 Myr.
   From the figure, we see that most of the observed sdB stars are located in the main
   sequence. However, we cannot confirm their core masses and their evolutionary
   stages yet, since a star with a different core mass in a
   different evolutionary stage is also likely in this region of
   the $T_{\rm eff}-{\rm log}(g)$ diagram (see Sect.3).
   For instance, an sdO star is possibly a main sequence star with a core
   mass higher than that of an sdB star, or in the stage of post main sequence of a sdB
   star.

\section{Methods}

   As we know, more than one
   evolutionary tracks pass through the same point on the $T_{\rm eff}- \rm log(g)$
   diagram. Thus, we cannot figure out the most probable one from them just
   by comparing the evolutionary tracks. To solve this problem, we intend to
   find another approach to deriving the masses of hot subdwarfs.

   Hot subdwarfs are core helium-burning stars with
   hydrogen envelopes that are too thin to sustain hydrogen burning;
   therefore, the luminosity of a hot subdwarf, $L$,
   depends upon the core mass, $M_{\rm c}$, and the age, $t$.
   Meanwhile, we know that
   $L \sim T_{\rm eff}^4 \cdot R^2 \sim M_{\rm c} \cdot \frac{T_{\rm eff}^4}{g}$,
   where $R$ is the radius. Thus, we have a relation as $f(M_{\rm c},t) \sim \log (\frac{T_{\rm eff}^4 }{g})$,
   where $T_{\rm eff}$ and $g$ are two fundamental parameters of hot subdwarfs from observations.
   Thus, we choose $\log (\frac{T_{\rm eff}^4}{g})$ as a basic
   parameter for further study of the masses of these objects.

   \begin{figure}
    \centering
     \includegraphics [angle=-90,scale=0.3]{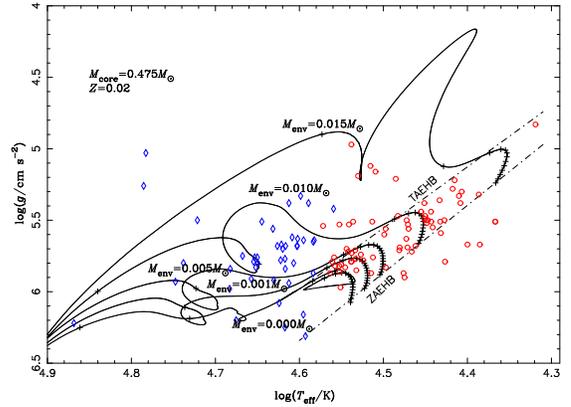}
     \caption{Evolutionary tracks of hot subdwarfs with a core of 0.475 $M_\odot$ and $Z$=0.02.
       The envelope masses are 0.000, 0.001, 0.005, 0.010, and 0.015 $M_\odot$. ZAEHB and TAEHB
       represent the onset and the end of main sequence, respectively (see the text for detail).
       The circles and diamonds are for sdB and sdO stars, respectively, from the ESO supernova Ia progenitor survey
      (SPY, Lisker et al. 2005, Stroeer et al. 2007).
       The age difference between two neighbouring crosses is 10 Myr.}
      \label{fig2}
   \end{figure}

   In general, stars evolves at different speeds in different evolutionary stages.
   If a star evolves slowly in a certain stage, it has more chance of being observed in this
   stage. We may then find out the most probable mass via studying the change rate of
   $\log (\frac{T_{\rm eff}^4}{g})$, as described in the following.
   We divide $\log (\frac{T_{\rm eff}^4
   }{g})$ into small intervals of 0.1 from 10.0 to 16.0,
   and calculate the evolutionary time, $\delta t$, of each
   model in these intervals. Thus, for each
   interval of $\log (\frac{T_{\rm eff}^4}{g})$, we may obtain a core mass,
   with which a hot subdwarf stays in it for the longest
   time amongst all the models (i.e. having the maximum value of $\delta
   t$), and has the most chance of being observed.
   This core mass is then the most probable mass, $M_{\rm p}$, in the observations, and is assumed to be the real mass of the subdwarf
   \footnote{Note that the mass given here is the core mass, which is a lower limit of the total mass of hot subdwarfs.}.
   We finally obtained a relation between $M_{\rm p}$ and $\log
   (\frac{T_{\rm eff}^4}{g})$, which is shown in Fig.3. From this
   relation we may simply know the mass of a hot subdwarf for given
   $T_{\rm eff}$ and $g$.

  \begin{figure}
    \centering
     \includegraphics [angle=0,scale=0.45]{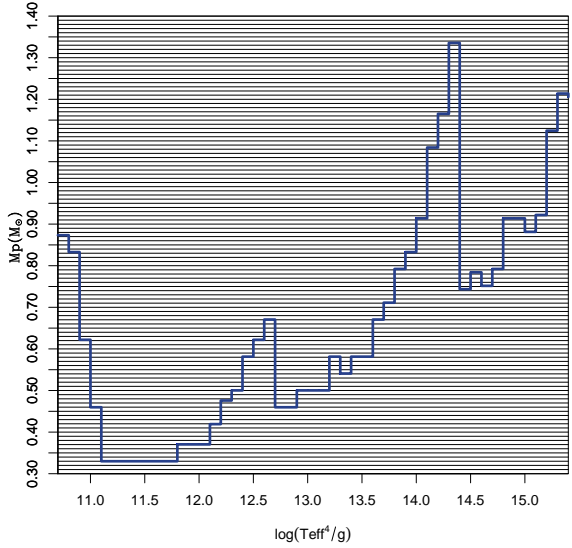}
     \caption{The $M_{\rm p}$-$\log
      (\frac{T_{\rm eff}^4}{g})$ relation. $M_{\rm p}$ is the most probable
      mass.}
      \label{fig3}
   \end{figure}

  Using the $M_{\rm p}$-$\log (\frac{T_{\rm eff}^4}{g})$ relation, we studied the
  masses of some observed hot subdwarfs, including 164 sdB stars and
  46 sdO stars. Some sdB stars are from the ESO supernova Ia progenitor
  survey (SPY, Lisker et al. 2005) and the other sdB stars are from the Hamburg quasar
  survey (HQS, Edelmann et al.2003). The sdO stars come from SPY (Stroeer et al. 2007).
  The mass distributions for sdB and sdO stars are shown in Figs.4 and
  5, respectively.

  The error on the mass, $\Delta M_{\rm p}$, depends on the
  parameter $\log(\frac{T_{\rm eff}^4}{g})$, and this dependence is shown in the upper
  panel of Fig.6, where $f$ ($ \equiv \frac{\Delta M_{\rm p}}{\Delta \log(T_{\rm eff}^4/g)
  }$) is derived from the $M_{\rm p}$-$\log
  (\frac{T_{\rm eff}^4}{g})$ relation. The $\log
  (\frac{T_{\rm eff}^4}{g})$ distributions of sdB and sdO stars from
  the SPY and the HQS are shown in the bottom panel of this figure. From this figure we
  see that $f$ is lower than 0.5 for most hot subdwarfs. To discuss the error on the mass obtained from our method,
  we should consider the error from both theory and observation.
  The $M_{\rm p}$-$\log
  (\frac{T_{\rm eff}^4}{g})$ relation here is derived from Pop I
  models ($Z$=0.02). Thus, the different metalicities result in
  a different $\log(\frac{T_{\rm
  eff}^4}{g})$ for the same mass. For example, $\Delta \log(\frac{T_{\rm
  eff}^4}{g})$ is about 0.04 dex for the models between $Z$ = 0.0001 and $Z$ = 0.06.
  For stars from the ESO supernova Ia progenitor survey, the observational errors
  for sdB stars are $\Delta T_{\rm eff}= 360K$ and $\Delta
  \log(\rm g)= 0.05 $ dex, and the observational errors for sdO stars are $\Delta \log(T_{\rm eff})= 0.011$ dex and $\Delta
  \log(\rm g)= 0.097 $ dex, respectively. These correspond to $\Delta \log(T_{\rm
  eff}^4/g)$ of 0.07  dex and 0.14 dex. Thus, the
  total errors, $\Delta \log(T_{\rm eff}^4/g)$ (theory plus observation), are 0.11 dex and 0.18 dex, respectively.
  According to the $f \sim \rm log(\frac{T_{\rm eff}^4}{g})$ relation shown in Fig.6,
  the errors on the masses are less than 0.055
  $M_\odot$ and 0.09 $M_\odot$ for most sdB and sdO stars, respectively. For sdB stars from the Hamburg Quasar
  Survey, we derived the mass errors for each star in a similar way. The results show that the majority of errors are
  in the 0.097 $M_\odot$ to 0.117 $M_\odot$ range(95 percent confidence interval), while the mean error is 0.107
  $M_\odot$.

\section{Results and discussion}

   \begin{figure}
    \centering
     \includegraphics [angle=0,scale=0.45]{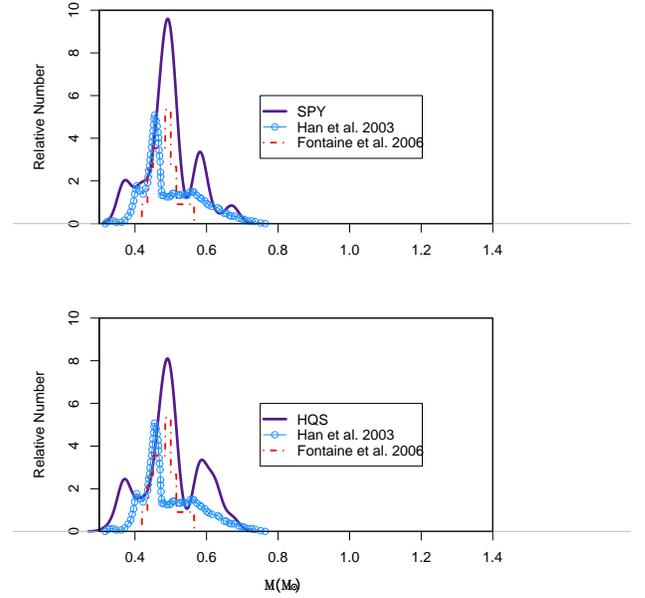}
     \caption{The mass distribution of sdB stars (derived from the $M_{\rm p}$-$\log (\frac{T_{\rm eff}^4}{g})$ relation)
      from the ESO supernova Ia progenitor survey (SPY, Lisker et al. 2005) and the Hamburg quasar
       survey (HQS, Edelmann et al.2003). As a comparison, we
      also show the mass distribution from the theoretical study of Han et al.
    (2003) and from asteroseismology
      (Fontaine et al. 2006).}
       \label{fig4}
   \end{figure}

  As shown in Fig.4, the sdB stars from SPY and HQS have the same
  mass distributions---the KS test shows that the two samples come from
  the same distribution at a high level of confidence($> 93.5\%$).
  Most of the sdB stars have masses ranging from 0.42 to
  0.54 $M_\odot$, and the mean mass is about 0.50 $M_\odot$,
  equal to that assumed in previous studies. As a comparison, we
  also show the mass distributions from the theoretical study of Han et al.
  (2003)
  \footnote{In the paper of Han et al. (2003), the mass distribution
  are for both sdB and sdO stars. We seperated these objects, thus the distribution
  presented here is different from that of Han et al.(2003).}
  and from asteroseismology (Fontaine et al. 2006) in Fig.4,
  from which we see that the distribution obtained in this paper
  matches the other two closely.

     \begin{figure}
    \centering
     \includegraphics [angle=0,scale=0.4]{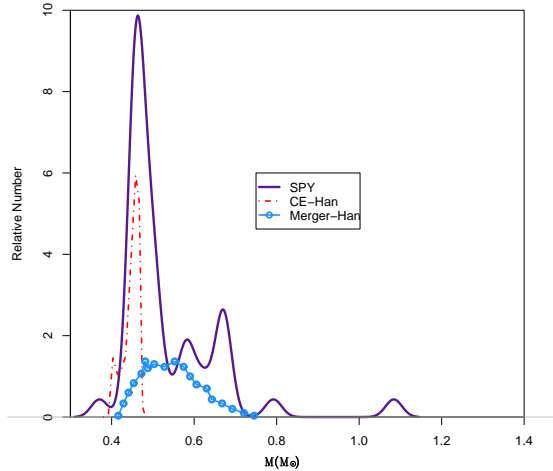}
     \caption{The mass distribution of sdO stars (derived from the $M_{\rm p}$-$\log (\frac{T_{\rm eff}^4}{g})$ relation)
      from the ESO supernova Ia progenitor survey (SPY, Lisker et al. 2005). The mass distributions from the CE and merger
       scenarios (Han et al. 2003$)^2$ are also shown in this figure.}
      \label{fig5}
     \end{figure}

  Figure 5 shows that most sdO stars are in 0.40 $\sim$
  0.55 $M_\odot$, although some sdO
  stars have much higher masses than these values. Since low-mass hot
  subdwarfs (e.g. less than $\sim$0.5 $M_\odot$) can not reach 40,000K
  (the minimum temperature for sdO stars) during the main
  sequence (see Fig.2 in Han et al. 2002), sdO stars with mass less than $\sim$0.5 $M_\odot$
  most likely evolve from sdB stars, i.e., these objects initially appear as
  sdB stars when they are on the main sequence, and as sdO stars when
  they evolved off the main sequence (post-EHB). The typical evolutionary time
  scales for hot subdwarfs on the MS and post-EHB are $\sim$ 160 and $\sim$ 20Myr,
  respectively. Thus, observationally the number of sdO stars (evolving from sdB stars) is
  expected to be about 12\% of that of sdB stars, and for SPY this
  equates to about 9 objects. They are likely He-deficient sdO stars.
  There are 13 He-deficient sdO stars in SPY, and 7 ones (likely evolve from sdB stars) of them with masses
  less than $\sim$0.5 $M_\odot$ in our study, similar to the 9 predicted by theory.

  To more easily compare with theoretical results, we presented the mass
  distributions of sdO stars obtained from Han et al. (2003)$^2$ in
  Fig.5, where the dot-dashed line is from CE ejection and the
  circle-line is from the merger of two He-WDs. We see that CE ejection can
  only account for sdO stars with masses lower than $\sim$0.5 $M_\odot$.
  The He-He merger mainly contributes to high-mass sdO stars, but it
  is also responsible for a small fraction of low-mass sdO stars.
  In particular, the fraction of very low-mass stars ($<
  0.4M_\odot$) is smaller among sdO stars than among sdB stars,
  as shown in Figs.4 and 5. This difference probably stems from
  very few sdB stars with masses lower than $0.4M_\odot$
  being hotter than 40,000K during their post-EHB stage.

  We know that a binary resulting from CE ejection generally has a
  short orbital period. Thus, if low-mass sdO stars are mainly from
  the CE ejection channel, then a large fraction of these objects should
  be observed in short orbital period binaries. Among 23 sdO stars with masses less than
  $\sim$0.5 $M_\odot$, (if they indeed evolve from sdB stars), we would expect a fraction
  of binaries similar to that of sdB stars, i.e. 39\% (Napiwotzki et al. 2004).
  However, only five sdO stars have been identified in binaries from
  SPY. The possible reasons for the low fraction are
  (a) the He-He merger contributes a small fraction of low-mass sdO
  stars, which are all singles; (b)low-mass sdO stars from
  CE ejection generally have lower mass companions (most likely M,
  see Fig.15 in Han et al. 2003), and may be identified as singles;
  (c) some other observational effects, such as random inclination of orbital planes,
  contribute to a decrease in detection efficiency.
  Moreover, in comparison to the merger channel, CE
  ejection, which may leave a thin hydrogen-rich envelope
  after the ejection, often produces hot subdwarfs with surface helium
  deficiency. Thus, if sdO stars are identified in binaries with short orbital periods, they are most likely to
  be helium deficient, since the merger of He-He WDs can only produce single helium-enriched sdO stars.
  The observations confirm this prediction; i.e., four of the five identified binary sdO
  stars have short orbital periods and are helium deficient. An
  extreme case from the CE ejection channel is that all of the common
  envelope has been ejected, leading to a naked He core (hot
  subdwarfs) and a companion. This may produce helium-enriched
  sdO stars with short orbital periods. Since this case is rare, the
  fraction of helium enriched sdO binaries among He-sdO stars (most of them are from the merger channel)
  is very low (4\% at most, Napiwotzki et al. 2004, see also Heber 2008).

    \begin{figure}
    \centering
     \includegraphics [angle=0,scale=0.35]{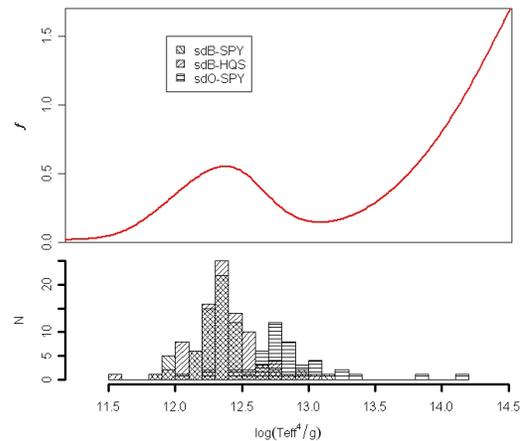}
     \caption{ The upper panel: the dependence of the mass errors on
      $\rm log(\frac{T_{\rm eff}^4}{g})$, derived from the $M_{\rm p}$-$\log (\frac{T_{\rm eff}^4}{g})$
      relation, where $f \equiv \frac{\Delta M_{\rm p}}{\Delta \log(T_{\rm eff}^4/g)
      }$. The bottom panel: the $\log
      (\frac{T_{\rm eff}^4}{g})$ distributions of sdB and sdO stars from
      the SPY and the HQS.}
      \label{fig6}
     \end{figure}

\section{Acknowledgements}
\tiny
      We thank an anonymous referee for his/her valuable comments
      that helped us to improve the paper. XZ thanks Dr. Richard
      Pokorny for improving the English language of the original
      manuscript. This work was in part supported by the Natural Science Foundation
      of China under Grant Nos.10821061,10603013, and
      2007CB815406, the Chinese Academy of Science under Grant
      Nos.06YQ011001, and the Yunnan National Science Foundation (Grant No.08YJ041001).

\end{document}